\documentclass[conference]{IEEEtran}
\IEEEoverridecommandlockouts
\usepackage{cite}
\usepackage[hidelinks]{hyperref}
\usepackage{amsmath,amssymb,amsfonts}
\usepackage{algorithmic}
\usepackage{graphicx}
\usepackage{placeins}
\usepackage{enumitem}
\usepackage{textcomp}
\usepackage[table]{xcolor}
\usepackage{threeparttable}
\usepackage{tikz}
\usepackage{booktabs}
\usepackage{pgfplots}
\usepackage{multirow}
\usepackage{tablefootnote}
\usepackage{pifont}
\pgfplotsset{compat=1.18}
\def\BibTeX{{\rm B\kern-.05em{\sc i\kern-.025em b}\kern-.08em
    T\kern-.1667em\lower.7ex\hbox{E}\kern-.125emX}}
\begin{document}

\title{From Business Problems to AI Solutions: Where Does Transformation Support Fail?}

\author{
\IEEEauthorblockN{Abir Trabelsi\IEEEauthorrefmark{1},
Imen Benzarti\IEEEauthorrefmark{1},
Hafedh Mili\IEEEauthorrefmark{2},
Darine Ameyed\IEEEauthorrefmark{3}}

\IEEEauthorblockA{\IEEEauthorrefmark{1}\textit{Software and IT Engineering},
\textit{École de technologie supérieure},
Montreal, Canada\\
abir.trabelsi.1@ens.etsmtl.ca, imen.benzarti@etsmtl.ca}

\IEEEauthorblockA{\IEEEauthorrefmark{2}\textit{Computer Science department},
\textit{Université du Québec à Montréal},
Montreal, Canada\\
mili.hafedh@uqam.ca}

\IEEEauthorblockA{\IEEEauthorrefmark{3}\textit{Computer Science and Mathematics},
\textit{Université du Québec à Chicoutimi},
Chicoutimi, Canada\\
darine\_ameyed@uqac.ca}
}

\maketitle

\begin{abstract}
Translating business problems into well-specified machine learning solutions is a prerequisite for successful AI systems, yet this upstream translation is still one of the least supported steps in existing methodologies. We conduct a structured narrative literature review of 18 approaches spanning requirements engineering (RE), machine learning (ML) project management, and automation. We organize these approaches into a taxonomy of four families and compare them across six input artifact categories, six output artifact categories, and a transformation framework of seven stages,  grounded in RE refinement theory and ML lifecycle process. Our study shows that most approaches list ML task or algorithm specification among their expected outputs, yet only four provide partial guidance for deriving it, and none provides systematic guidance. We characterize this gap as the Analytics Translation Problem (ATP) and derive five research recommendations addressing multi-formulation exploration, task derivation guidance, constraint-algorithm filtering, probabilistic traceability, and data-triggered revision.  
These findings outline a focused research agenda for the translation step largely left to practitioner intuition.
\end{abstract}


\section{Introduction}
\label{sec:introduction}

AI projects fail at rates estimated to be twice those of conventional IT projects~\cite{Ryseff2024}. Misalignment between the business problem stakeholders intend to solve and the technical formulation delivered to data science teams has been identified as a recurring root cause through interviews with 65 practitioners~\cite{Ryseff2024}. When stakeholders express a need such as "reduce customer churn" or "detect fraudulent transactions", data science teams must determine what machine learning (ML) task to formulate (e.g., classification, survival analysis, clustering), which data to collect, and which metrics operationalize success. This upstream translation from business problem (BP) to ML solution specification (MLS) is still mostly ad hoc, reliant on individual expertise, and weakly supported by existing methodologies. As a result, teams may construct technically sound models that do not address the intended business objective, or spend extended periods iterating on problem formulation before modeling begins~\cite{REF4, REF61}.

This challenge is, at its core, a \emph{requirements engineering} (RE) problem. The steps of eliciting stakeholder needs, specifying system requirements, ensuring problem-solution alignment, and validating that solutions address original needs are central concerns of RE research~\cite{REF20, REF21, REF22}. Yet, the BP$\rightarrow$MLS transformation differs from traditional RE in a fundamental way. Such transformation requires \emph{cross-domain derivation}, translating business characterizations expressed in the language of goals, decisions, and constraints into statistical and algorithmic constructs (ML paradigms, task types, evaluation metrics) that have no semantic correspondence in the original problem statement. This is what sets the problem apart from classical goal-to-requirement refinement~\cite{REF20, REF21}, and it goes a long way toward explaining why neither traditional RE methods nor data science process models have managed to support it systematically.

Recent RE research addressed AI-specific concerns, including requirements for ML-enabled systems~\cite{REF59, REF60}, non-functional requirements for ML~\cite{REF62}, and development challenges for AI-intense systems~\cite{REF61}. In parallel, the data science community has developed lifecycle models such as CRISP-DM~\cite{REF6, REF7}, MLOps frameworks~\cite{REF19}, RE-inspired approaches (GR4ML~\cite{REF51}, RE4HCAI~\cite{REF67}), and automation-oriented approaches that address algorithm selection downstream~\cite{REF18}. Still, these efforts are fragmented and, as we demonstrate, provide little systematic guidance for deriving the ML task from business characterizations.

To address this fragmentation, we conduct a narrative literature review examining how 18 existing approaches, spanning requirements engineering, data science project management, and automation, support the BP$\rightarrow$MLS transformation. We analyze these approaches through a common analytical view (input/output artifacts, transformation mechanisms, and stage coverage) to identify where support exists, where it fails, and what a transformation framework would need to provide.

In this paper, we first propose a four-family taxonomy that organizes $18$ BP$\rightarrow$AI transformation approaches from fragmented research streams into a coherent landscape (Section~\ref{sec:taxonomy}). Second, we provide a comparative artifact- and stage-based analysis revealing the limitations in existing transformation mechanisms (Section~\ref{sec:comparative}). Third, we characterize the Analytics Translation Problem (ATP) as the absence of explicit, traceable, and operationalizable mappings between business characterizations and ML formulations (Section~\ref{sec:synthesis}). Fourth, we derive five evidence-based recommendations (R1--R5) specifying what future transformation frameworks should provide to address the gaps (Section~\ref{sec:synthesis}). This work positions the BP→MLS transformation as a distinct RE research problem and aims to move the translation step from practitioner intuition toward principled, traceable refinement mechanisms.

\section{Background}

\label{sec:background}

This section defines key terms:

\textbf{Business Problem (BP).} We define a \textbf{BP} as a measurable gap an organization seeks to close, characterized by four elements: \emph{strategic objectives}, \emph{decision needs} that stakeholders want to support with data-driven insights, \emph{performance gaps} between current and target indicators, and \emph{constraints} (budget, timeline, regulatory, ethical, technical)~\cite{REF8, REF20, REF10, REF9}. BPs are multi-causal and articulated at varying abstraction levels,  consequently, they require progressive refinement to achieve addressable formulations~\cite{saleh2024matching}.

\textbf{ML Solution Specification (MLS).} We define an MLS as the specification bridging business requirements and ML implementation. An MLS comprises the \emph{ML paradigm} (supervised, unsupervised, reinforcement, or hybrid), the \emph{task type} (classification, regression, clustering, anomaly detection, etc.), \emph{data requirements}, \emph{constraint operationalization} (e.g., ``explainable'' $\rightarrow$ interpretability requirements), \emph{evaluation criteria} aligned with business thresholds, and \emph{operational context}~\cite{REF2, REF4}.

\textbf{Transformation.} We define transformation as deriving a specified MLS from a characterized BP such that the MLS addresses the BP, the conceptual link is explicit and traceable, and stakeholders can validate that the intent has been captured.

\section{Review Methodology}
\label{sec:method}

This section presents the research design of our review, covering the research questions, search strategy, data extraction, and analysis framework.

\subsection{Research Questions}
\label{subsec:rqs}

Our review is guided by four research questions that progressively characterize the BP$\rightarrow$MLS transformation landscape:
\begin{itemize}
    \item \textbf{RQ1:} How can existing BP$\rightarrow$MLS approaches 
          be organized by methodological orientation?
    \item \textbf{RQ2:} What problem-space and solution-space artifacts 
          do these approaches consume and produce?
    \item \textbf{RQ3:} What types of transformation mechanisms do 
          these approaches employ?
    \item \textbf{RQ4:} Which transformation stages receive systematic 
          support, and where do approaches fail to specify 
          \emph{how} to derive outputs from inputs?
\end{itemize}

\subsection{Review Design and Search Strategy}
\label{subsec:search}

We adopt a narrative literature review methodology~\cite{green2006narrative}, appropriate for synthesizing diverse research streams and identifying conceptual gaps across heterogeneous approaches. A systematic review~\cite{kitchenham2007guidelines} was not feasible here because our corpus spans multiple fields, requirements engineering, data science, and automation that use incompatible terminologies and contribution types, making uniform quality scoring and quantitative aggregation impractical. The structured extraction schema and classification procedures we employ are intended to enhance transparency and support the replicability of our analysis, rather than to achieve the inferential validity of an SLR. We made the complete extraction data publicly available.

We searched five digital libraries (IEEE Xplore, ACM Digital Library, Springer Link, ScienceDirect, arXiv) and targeted venues spanning RE and data science communities. 
The search combined terms across three domains problem ("business problem", "business goal", "problem formulation"), solution ("machine learning", artificial intelligence, "data science"), and transformation ("requirements engineering" ,"specification", "process model") using Boolean operators.

Our four-stage selection process yielded 129 initial articles (Stage~1), reduced to 101 after title/abstract screening and deduplication (Stage~2), then to 36 candidates after full-text review for concrete BP$\rightarrow$
AI transformation approaches (Stage~3). Quality and scope filtering produced the final corpus of 18 approaches (Stage~4),  the 18 excluded candidates addressed only a single downstream activity, provided insufficient methodological detail, or were superseded by a more comprehensive version already retained.


Inclusion required that a paper (i)~propose a concrete approach, (ii)~address transformation from business needs to AI/ML specifications, (iii)~provide sufficient methodological detail, and (iv)~appear in a peer-reviewed venue. We excluded pure algorithm papers, deployment-only papers without upstream formulation, position papers without a concrete methodology, and superseded versions. When multiple papers described the same approach (e.g., GR4ML~\cite{REF51, REF52, REF53, REF54, REF55, REF56}), we selected the most comprehensive version. Theoretical saturation served as the stopping rule. We continued the search until new articles no longer introduced novel approach types or mechanisms. Specifically, the last five articles reviewed (chronologically) mapped entirely onto existing families and mechanism types, introducing no new coverage patterns.

\subsection{Data Extraction}
\label{subsec:extraction}

We developed a structured schema comprising 19 fields organized into five categories, as shown in Table~\ref{tab:extraction-schema}.

\begin{table}[htbp]
\centering
\caption{Data Extraction Schema (19 Fields)}
\label{tab:extraction-schema}
\scriptsize
\renewcommand{\arraystretch}{1.15}
\begin{tabular}{lp{5.5cm}}
\toprule
\textbf{Field} & \textbf{Description} \\
ID & Unique identifier (1--18) \\
Title & Full article title \\
Publication Type & Journal / Conference / Other \\
Venue & Publication venue name \\
Year & Publication year \\
Solution Family & RE4AI / Data-Centric / PM / Automation \\
Transformation? & Yes / No / Partially \\
Problem Category & Business problem / User need / Requirement \\
Solution Category & AI / ML / DL / Data Mining / Data Science \\
Models Problem? & Yes / No / Partially \\
Problem Properties & Free text: what problem aspects are modeled \\
Models AI Solution? & Yes / No / Partially \\
Solution Properties & Free text: what solution aspects are modeled \\
Solution Type & Process / Model / Framework / Catalogue / Ontology \\
Steps Description & Free text: methodology steps or phases \\
Validated? & Yes / No \\
Validation Method & Industrial case / Academic case / Expert evaluation / None \\
Limitations & Free text: explicitly stated limitations \\
Related Articles & References to related or prior work \\
\bottomrule
\end{tabular}
\end{table}

The first author performed the extraction. The second author verified and validated the extraction through discussions. 

\subsection{Analysis Framework}
\label{subsec:analysis-framework}

We grouped the 18 approaches into four families by methodological orientation: RE for AI, Data-Centric RE, Project Management Processes, and Automation-Oriented. Each approach was assigned to the family matching its primary contribution, when one straddled two families, we went with its dominant focus.

Coverage was assessed against a seven-stage transformation framework (S1--S7), whose stages are grounded in established RE and data science process theory (see Section~\ref{sec:comparative}). Each approach was rated per stage as strong ($\bullet\bullet\bullet$, explicit guidance with concrete mechanisms), partial ($\bullet\bullet\circ$, mentioned but not operationalized), or minimal ($\bullet\circ\circ$, absent or only a passing reference).

To assess rating reliability, the second author independently rated six approaches—spanning all four families, across all seven stages using Table~\ref{tab:criteria}, without seeing the first author's ratings. Weighted Cohen's kappa came to $\kappa = 0.83$ (linear weighting), which indicates excellent agreement. On S3, the stage most central to our findings, agreement was perfect. The six disagreements all fell between adjacent categories; a calibration session resolved them, with two ratings changed. For gap identification, we worked bottom-up: we started from the limitations each paper's own authors stated, looked for recurring weakness patterns in the coverage data, and then synthesized these into capability gaps shared by the majority of approaches.

\section{Results I: Taxonomy of BP-to-AI Transformation Approaches}

\label{sec:taxonomy}

This section addresses \textbf{RQ1} by presenting a taxonomy of 18 identified approaches organized into four families based on methodological orientation. Table~\ref{tab:approach-inventory} presents the complete inventory of 18 approaches with their classification, solution type, and validation status.

\begin{table*}[h]

\centering

\caption{Inventory of BP$\rightarrow$AI Transformation Approaches (n=18) }

\label{tab:approach-inventory}

\scriptsize

\begin{tabular}{clllllc}

\toprule

\textbf{ID} & \textbf{Approach} & \textbf{Year} & \textbf{Family} & \textbf{Solution Type} & \textbf{Solution Category} & \textbf{Validated} \\

\midrule


1 & RE4HCAI~\cite{REF67} & 2023 & RE4AI & Conceptual Framework & AI & Industrial \\

4 & Catalogue of Concerns~\cite{REF71} & 2022 & RE4AI & Catalogue + Template & ML & Focus Group \\

6 & ML Snapshot Framework~\cite{REF82} & 2022 & RE4AI & Framework & ML & Academic \\

7 & GR4ML~\cite{REF51} & 2021 & RE4AI & Process + Catalogues & ML & Industrial \\

8 & RM4ML~\cite{REF96} & 2025 & RE4AI & Process & ML & Industrial \\

\midrule


2 & Extended AMDiRE~\cite{REF70} & 2021 & Data-Centric & Model & AI/ML & Not Validated \\

3 & MLC Specification~\cite{REF69} & 2019 & Data-Centric & Ontology & ML & Not Validated \\

5 & Multi-layered Framework~\cite{REF81} & 2023 & Data-Centric & Process + Template & ML & Academic \\

\midrule






11 & CRISP-DM~\cite{REF6, REF7} & 2006 & PM & Process & Data Mining & Industrial \\

12 & DST~\cite{REF5} & 2021 & PM & Model & Data Science & Industrial \\

13 & AI Lifecycle Revisions~\cite{REF15} & 2021 & PM & Model & DS/ML & Industrial \\

14 & CRISP-ML(Q)~\cite{REF17} & 2021 & PM & Process & DS/ML & Not Validated \\

15 & Interpretability ML~\cite{REF107} & 2021 & PM & Conceptual Framework & DS/ML & Industrial \\

9 & AI Use Case Development~\cite{REF108} & 2020 & PM & Process & AI & Expert Interviews \\

10 & BDDS~\cite{REF4} & 2024 & PM & Process & AI & Industrial \\

16 & MAISTRO~\cite{REF106} & 2025 & PM & Process & AI & Academic \\

\midrule


17 & MLOps Architecture~\cite{REF19} & 2023 & Automation & Process & ML & Expert Interviews \\

18 & SmartML~\cite{REF18} & 2019 & Automation & Framework & ML & Academic \\

\bottomrule

\end{tabular}

\begin{tablenotes}

\scriptsize

\item Validation categories: Industrial = real-world case study,  Academic = fictitious/simulated case,   Focus Group/Expert Interviews = expert evaluation,  Illustrative = example-based demonstration.

\end{tablenotes}

\end{table*}

\subsection{Family A: Requirements Engineering for AI (RE4AI)}

\label{subsec:re4ai}

This family comprises five approaches extending traditional RE practices with AI-specific catalogues, models, or frameworks. We cluster them into two categories:
The first category comprises approaches focused on specification that propose catalogs, perspectives, and UML models for organizing AI requirements. However, they lack transformation mechanisms. RE4HCAI~\cite{REF67} captures human-centered concerns through three layers,  the Catalogue of Concerns~\cite{REF71} organizes five perspectives (ML Objectives, UX, Infrastructure, Model, Data),  and RM4ML~\cite{REF96} models ML components via UML.
The second category consists of approaches that explicitly attempt a connection from business to technical. GR4ML~\cite{REF51} extends goal-oriented RE through three views (Business, Analytics Design, Data Preparation), which constitute the most explicit bridging attempt in this family, but industrial validation revealed that the Business$\rightarrow$Analytics transition remains underspecified. ML Snapshot~\cite{REF82} uses metamorphic relations for behavioral specifications but has been narrowly validated.

\subsection{Family B: Data-Centric Requirements Engineering}

\label{subsec:datacentric}

Three approaches prioritize data requirements as the bridge between business problems and ML solutions. Extended AMDiRE~\cite{REF70} augments a traditional three-layer artifact model with a data-centric layer covering datasets, algorithms, and metrics. MLC Specification~\cite{REF69} introduces a web-based ontology to support unambiguous domain conceptualization and dataset benchmarking. The Multi-layered Framework~\cite{REF81} targets safety-critical ML through Problem, Data, and Evidence layers and is validated on pedestrian detection datasets. Despite their differences, Extended AMDiRE~\cite{REF70} lacks empirical validation, and MLC Specification~\cite{REF69} omits task formulation.




\subsection{Family C: Project Management processes for AI/DS}

\label{subsec:project_management_processes}

Eight approaches provide lifecycle models that define when activities occur. We cluster them into two categories:

The first category encompass approaches \textit{focusing on Lifecycle Management}. These approaches emphasize governance, quality assurance, and operational concerns. CRISP-DM~\cite{REF6, REF7}, the most widely adopted model (6 phases), provides minimal guidance on Business Understanding despite its industrial ubiquity. DST~\cite{REF5} extends CRISP-DM with exploratory activities,  AI Lifecycle revised models~\cite{REF15} add Go/No-Go gates and monitoring,  CRISP-ML(Q)~\cite{REF17} merges Business and Data Understanding with quality checkpoints,  CRISP-ML framework~\cite{REF107} emphasizes interpretability throughout 7 phases,  MAISTRO~\cite{REF106} focuses on ethics and bias.

The second category encompass \textit{formulation approaches}. These approaches explicitly address task formulation. BDDS~\cite{REF4} addresses data requirements by deriving the "right data" from the "right question" identified through root cause analysis and supports feasibility assessment through its DIEM framework. The framework classifies business situations across four quadrants based on the understanding of relevant system dependencies and data availability, prescribing AI methods (Quadrant 1) when dependencies are opaque but data are plentiful,  but, the guidance is still qualitative without specific algorithm recommendations.
AI Use Cases~\cite{REF108} provides explicit problem-to-AI capability matching through cognitive function mapping and problem–solution matrices, validated via expert interviews but lacking real-project validation. However, this mapping continues to be at the level of abstract capabilities: knowing that a problem requires \textit{prediction} does not indicate whether linear regression, gradient boosting, or a recurrent neural network is appropriate, nor how data or operational constraints should guide the choice.

\subsection{Family D: Automation-Oriented Approaches }

\label{subsec:automation}

We identify two approaches that focus on downstream automation, explicitly assuming upstream formulation is complete. MLOps Architecture~\cite{REF19} defines a reference architecture (Feature Engineering, Experimentation, Automated Pipeline) and addresses operational challenges but does not elaborate on Project Initiation and assumes business understanding is already established. SmartML~\cite{REF18} tackles algorithm selection and hyperparameter tuning through meta-learning—matching dataset meta-features against a knowledge base built from 50 OpenML, UCI, and Kaggle datasets. SmartML handles algorithm choice once the ML task is fixed, not based on the business problem.

\subsection{Cross-Family Analysis}

\label{subsec:synthesis}

Empirical grounding varies across families. Eight approaches report industrial validation, four rely on academic cases, and three lack validation entirely.

All 18 approaches provide partial transformation support. Each family addresses different transformation aspects requirements organization (Family A), data specification (Family B), lifecycle management (Family C), and post-formulation optimization (Family D). The translation from BP to MLS remains largely implicit across all families.

\section{Results II: Comparative Analysis of Transformation Support}
\label{sec:comparative}

This section addresses \textbf{RQ2} (What artifacts are produced?), \textbf{RQ3} (What mechanism types do approaches employ?), and \textbf{RQ4} (What coverage and gaps exist?) through a systematic comparative analysis of the 18 identified approaches. We examined input/output artifacts, transformation mechanisms, and coverage completeness using a common analytical lens.

\subsection{Input and Output Artifacts (RQ2)}
\label{subsec:artifacts}

We analyze the artifacts each approach consume as inputs and produce as outputs to understand the transformation support each approach provides, Table~\ref{tab:input_artifacts} summarizes problem-space (input) modeling and Table~\ref{tab:output_artifacts}  summarizes solution-space (output) modeling.

\subsubsection{Problem Space Modeling (Input Artifacts)}

By synthesizing across approaches, we identify six categories of input artifacts that characterize problem-space modeling, as shown in Table~\ref{tab:input_artifacts}. 
Most approaches capture strategic objectives and domain context, with formalization ranging from informal descriptions~\cite{REF6,REF5} to structured goal hierarchies~\cite{REF51,REF4}. Fewer than half model stakeholder specifications or data characteristics as explicit input artifacts.

\begin{table*}[h]
\centering
\caption{Problem Space Input Artifacts: Six Categories with Coverage and Sub-Element References (n=18)}
\label{tab:input_artifacts}
\scriptsize
\setlength{\tabcolsep}{4pt}
\begin{tabular}{|p{3cm}|p{1.5cm}|p{12.5cm}|}
\hline
\rowcolor{blue!20}
\textbf{Input Artifact Category} & \textbf{Coverage} & \textbf{Sub-Elements with References} \\
\hline
\rowcolor{gray!10}
\textbf{Strategic Objectives} & 
67\%  (12/18) & 
Business goals~\cite{REF70, REF71, REF81, REF51, REF4, REF6, REF5, REF15, REF17, REF107, REF106},  
Success criteria~\cite{REF17, REF106},  
Performance gaps~\cite{REF82, REF4} \\
\hline
\textbf{Constraints and Rules} & 
50\%  (9/18) & 
Constraints~\cite{REF67, REF70, REF82, REF96, REF17, REF107},  
Regulatory/Legal~\cite{REF108, REF15},  
Ethical criteria~\cite{REF108, REF17, REF106},  
Quality attributes~\cite{REF70, REF96} \\
\hline
\rowcolor{gray!10}
\textbf{Domain Context} & 
56\%  (10/18) & 
Business processes~\cite{REF70},  
Operational environment~\cite{REF81, REF82, REF96, REF108, REF17, REF107, REF106},  
Domain aspects~\cite{REF70, REF71, REF81, REF69} \\
\hline
\textbf{Data Characteristics} & 
56\%  (10/18) & 
Data availability~\cite{REF69, REF15, REF17, REF107, REF19},  
Data quality~\cite{REF69, REF15, REF17, REF107, REF19},  
Features~\cite{REF67, REF69, REF81, REF82, REF108, REF18},  
Datasets~\cite{REF69, REF15, REF19} \\
\hline
\rowcolor{gray!10}
\textbf{Stakeholder Specification} & 
44\%  (8/18) & 
User identification~\cite{REF67, REF71, REF51},  
Stakeholder maps~\cite{REF70, REF15, REF107,  REF106},  
Actors/Roles~\cite{REF51, REF96} \\
\hline
\textbf{Feasibility Assessment} & 
50\%  (9/18) & 
Go/No-Go criteria~\cite{REF67, REF108, REF15, REF17, REF96, REF4},  
Risk analysis~\cite{REF81, REF15, REF107, REF106},  
Viability evaluation~\cite{REF107} \\
\hline
\end{tabular}
\end{table*}

\subsubsection{Solution Space Modeling (Output Artifacts)}
 We identify six output categories in Table~\ref{tab:output_artifacts}. Data requirements~\cite{REF70,REF81,REF6,REF5,REF17,REF106} are most commonly modeled, reflecting the data-centric nature of ML. Recent approaches~\cite{REF17,REF106,REF70} increasingly emphasize AI-specific quality attributes beyond traditional NFRs. Evaluation criteria are often underspecified, with approaches listing metrics without defining thresholds or business-aligned acceptance criteria. PM-driven~\cite{REF6,REF17,REF106} and automation-oriented approaches~\cite{REF19} address deployment, while RE4AI approaches~\cite{REF67,REF71,REF51,REF96,REF82} largely stop at specification. Only four approaches~\cite{REF51,REF70,REF108,REF18} explicitly address algorithm or task selection. These four receive partial rather than strong S3 ratings because each addresses only a fragment of ML task formulation. SmartML automates algorithm selection \emph{within} a predefined task (e.g., selecting among classifiers) but cannot determine \emph{whether} classification is the appropriate task. AI Use Cases maps cognitive functions to problem types but yields coarse categories (e.g., ``prediction'') without specifying task type, paradigm, or data constraints. GR4ML decomposes goals into question goals but provides no derivation rules from questions to ML tasks. BDDS uses root-cause analysis to frame questions but is still qualitative. Strong S3 would require guiding practitioners from a business question to a specified ML task, none of the four does this.

\begin{table*}[h]
\centering
\caption{Solution Space Output Artifacts: Six Categories with Coverage and Sub-Element References (n=18) }
\label{tab:output_artifacts}
\scriptsize
\setlength{\tabcolsep}{4pt}
\begin{tabular}{|p{3.5cm}|p{1.5cm}|p{12cm}|}
\hline
\rowcolor{blue!20}
\textbf{Output Artifact Category} & \textbf{Coverage} & \textbf{Sub-Elements with References} \\
\hline
\rowcolor{gray!10}
\textbf{Data Requirements} & 
94\%  (17/18) & 
Dataset specifications~\cite{REF70, REF81, REF96},  
Data quality criteria~\cite{REF67, REF69, REF71, REF81, REF82, REF4, REF106},  
Feature definitions~\cite{REF70, REF51, REF15, REF19},  
Data preparation~\cite{REF67, REF51, REF108, REF4, REF6, REF5, REF15, REF17, REF107, REF106, REF19} \\
\hline
\textbf{Algorithm/Task Specification} & 
83\%  (15/18) & 
ML task formulation~\cite{REF51},  
Algorithm selection~\cite{REF67, REF70, REF71, REF51, REF96, REF6, REF5, REF15, REF17, REF107, REF106, REF19, REF18},  
AI solution types~\cite{REF108, REF4},  
Cognitive function mapping~\cite{REF108} \\
\hline
\rowcolor{gray!10}
\textbf{Evaluation Criteria} & 
78\%  (14/18) & 
Performance metrics~\cite{REF67, REF70, REF71, REF51, REF96, REF6, REF15, REF5, REF17, REF107, REF106, REF19},  
Validation methods~\cite{REF6, REF5},  
Acceptance criteria~\cite{REF71, REF81, REF82, REF17},  
Bias detection~\cite{REF15, REF107, REF106} \\
\hline
\textbf{Quality Attributes / NFRs} & 
67\%  (12/18) & 
Explainability~\cite{REF67, REF70, REF71},  
Fairness~\cite{REF70, REF96, REF15, REF17, REF106},  
Robustness~\cite{REF81, REF82, REF51},  
Trustworthiness~\cite{REF70, REF108},  
Interpretability~\cite{REF96, REF15, REF107} \\
\hline
\rowcolor{gray!10}
\textbf{Deployment Considerations} & 
56\%  (10/18) & 
Operational integration~\cite{REF108, REF6, REF5, REF15, REF17, REF107, REF106, REF19},  
Infrastructure~\cite{REF71},  
Monitoring~\cite{REF15, REF17, REF107, REF106, REF19},  
Maintenance~\cite{REF71, REF108, REF17},  
Feedback mechanisms~\cite{REF67} \\
\hline

\textbf{Traceability Links} & 
33\%  (6/18) & 
Goal-to-algorithm mapping~\cite{REF51, REF4, REF5, REF17},  
Requirements traceability~\cite{REF70},  
Layer connections~\cite{REF81} \\
\hline
\end{tabular}
\end{table*}

\subsection{Transformation Mechanisms and Explicitness (RQ3)}
\label{subsec:mechanisms}

We classify each approach by its primary mechanism type (Table~\ref{tab:approach-inventory}, column 5). Process/Methodology dominates (9/18), followed by Framework (4/18), Model (3/18), Catalogue (1/18), and Ontology (1/18). Several approaches~\cite{REF51, REF71, REF81} combine multiple types. Regardless of mechanism, all share a limitation in this respect: processes define when transformation activities occur (phase sequencing) but do not provide guidance about how to execute decisions within phases,  models and catalogues specify what to capture but not how to derive ML formulations from the captured information and automation tools optimize within a predefined task without guiding task selection.

\subsection{Coverage and Completeness (RQ4)}
\label{subsec:coverage}

This subsection examines the completeness of each approach’s transformation support by mapping coverage against a stage framework to identify where gaps exist.

\subsubsection{Transformation Stage Framework}

To assess coverage systematically, we define seven stages that a complete BP$\rightarrow$MLS transformation should address. Table~\ref{tab:stage-framework} provides definitions for each stage.

\begin{table*}[htbp]
\centering
\caption{BP$\rightarrow$MLS Transformation Stage Framework}
\label{tab:stage-framework}
\scriptsize
\renewcommand{\arraystretch}{1.15}
\begin{tabular}{p{0.4cm} p{3.5cm} p{8cm} p{4.5cm}}
\toprule
\textbf{Stage} & \textbf{Name} & \textbf{Description} & \textbf{External Grounding} \\
\midrule
S1 & Problem Framing & Identify business goals, stakeholders, performance gaps, and decision needs & i* framework~\cite{REF20},  BABOK~\cite{REF8} \\
S2 & Decision/ Question Formulation & Translate goals into specific questions or decisions the system should support & GQM paradigm~\cite{basili1994},  goal refinement~\cite{REF21} \\
S3 & ML Task Formulation & Derive ML paradigm and task type (classification, regression, etc.) & ML taxonomy~\cite{bishop2006},  task mapping~\cite{domingos2012} \\
S4 & Data Requirements & Specify data needs, quality criteria, availability, and feasibility & ISO/IEC 25012~\cite{iso25012},  Datasheets~\cite{gebru2021} \\
S5 & Constraints Integration & Incorporate NFRs, ethical and regulatory constraints into specifications & ISO/IEC 25010~\cite{iso25010},  NFRs for ML~\cite{REF62} \\
S6 & Evaluation Criteria & Define metrics, thresholds, and validation approach aligned with business goals & Sokolova \& Lapalme~\cite{sokolova2009},  Flach~\cite{flach2012} \\
S7 & Deployment Assumptions & Specify operational context, monitoring, and maintenance expectations & ML technical debt~\cite{sculley2015},  MLOps~\cite{REF19} \\
\bottomrule
\end{tabular}
\end{table*}

Stages S1--S2 constitute the problem space,  S4--S7 constitute the solution space,  S3 represents the translation point where business characterizations must be converted into ML formulations. Each stage is grounded in sources external to the reviewed corpus (Table~\ref{tab:stage-framework}, column~4), ensuring that the identified gaps reflect the absences in existing practice rather than artifacts of our framework design.

\subsubsection{Stage Coverage Analysis}

Table~\ref{tab:coverage-matrix} presents our stage-by-stage assessment using the criteria defined in Table~\ref{tab:criteria}. Ratings are grounded in documented capabilities and the authors' stated limitations.

\begin{table}[h]
\centering
\caption{Stage Coverage Rating Criteria (Abbreviated)}
\label{tab:criteria}
\scriptsize
\begin{tabular}{cp{7cm}}
\toprule
\textbf{Stage} & \textbf{Strong ($\bullet\bullet\bullet$) Coverage Criteria} \\
\midrule
S1 & 3+ problem elements (goals, stakeholders, constraints) 
     OR a dedicated problem-modeling phase \\
S2 & Explicit ``decision goals'' OR ``question goals'' artifacts \\
S3 & Algorithm-selection guidance OR paradigm mapping 
     OR task-derivation rules \\
S4 & A dedicated data layer OR 3+ data elements 
     (datasets, quality, features) \\
S5 & 3+ AI-specific attributes (explainability, fairness, 
     ethics, robustness) \\
S6 & Metrics AND business-aligned thresholds \\
S7 & Deployment + monitoring/maintenance phases \\
\bottomrule
\end{tabular}
\begin{tablenotes}\scriptsize
\item Partial ($\bullet\bullet\circ$): 1 or 2 elements OR mention without detail. 
Minimal ($\bullet\circ\circ$): No coverage OR explicit absence stated.
Full criteria and evidence mapping are available in the online repository.
\end{tablenotes}
\end{table}

\begin{table*}[h]
\centering
\caption{Stage Coverage Matrix: BP$\rightarrow$MLS Transformation (n=18) }
\label{tab:coverage-matrix}
\scriptsize
\renewcommand{\arraystretch}{1.1}
\begin{tabular}{cl|ccccccc|c}
\toprule
\textbf{ID} & \textbf{Approach} & \textbf{S1} & \textbf{S2} & \textbf{S3} & \textbf{S4} & \textbf{S5} & \textbf{S6} & \textbf{S7} & \textbf{Complete?} \\
& & \tiny{Characterize} & \tiny{Formulate} & \tiny{Select} & \tiny{Specify} & \tiny{Define} & \tiny{Establish} & \tiny{Plan} \\
& & \tiny{Problem} & \tiny{Decisions} & \tiny{ML Task} & \tiny{Data Reqs} & \tiny{Constraints} & \tiny{Evaluation} & \tiny{Deployment} \\
\midrule
\multicolumn{10}{l}{\emph{Family A: RE for AI (RE4AI)}} \\
1 & RE4HCAI~\cite{REF67} & $\bullet\bullet\circ$ & $\bullet\circ\circ$ & $\bullet\circ\circ$ & $\bullet\bullet\bullet$ & $\bullet\bullet\circ$ & $\bullet\bullet\circ$ & $\bullet\circ\circ$ & No \\
4 & Catalogue~\cite{REF71} & $\bullet\bullet\circ$ & $\bullet\bullet\circ$ & $\bullet\circ\circ$ & $\bullet\bullet\circ$ & $\bullet\bullet\circ$ & $\bullet\bullet\circ$ & $\bullet\bullet\circ$ & No \\
6 & ML Snapshot~\cite{REF82} & $\bullet\bullet\circ$ & $\bullet\circ\circ$ & $\bullet\circ\circ$ & $\bullet\bullet\circ$ & $\bullet\bullet\circ$ & $\bullet\bullet\circ$ & $\bullet\circ\circ$ & No \\
7 & GR4ML~\cite{REF51} & $\bullet\bullet\bullet$ & $\bullet\bullet\bullet$ & $\bullet\bullet\circ$ & $\bullet\bullet\bullet$ & $\bullet\bullet\circ$ & $\bullet\bullet\circ$ & $\bullet\circ\circ$ & No \\
8 & RM4ML~\cite{REF96} & $\bullet\bullet\circ$ & $\bullet\circ\circ$ & $\bullet\circ\circ$ & $\bullet\bullet\circ$ & $\bullet\bullet\circ$ & $\bullet\bullet\circ$ & $\bullet\circ\circ$ & No \\
\midrule
\multicolumn{10}{l}{\emph{Family B: Data-Centric RE}} \\
2 & Ext. AMDiRE~\cite{REF70} & $\bullet\bullet\bullet$ & $\bullet\circ\circ$ & $\bullet\circ\circ$ & $\bullet\bullet\circ$ & $\bullet\bullet\circ$ & $\bullet\circ\circ$ & $\bullet\circ\circ$ & No \\
3 & MLC Spec.~\cite{REF69} & $\bullet\circ\circ$ & $\bullet\circ\circ$ & $\bullet\circ\circ$ & $\bullet\bullet\circ$ & $\bullet\circ\circ$ & $\bullet\circ\circ$ & $\bullet\circ\circ$ & No \\
5 & Multi-layer~\cite{REF81} & $\bullet\bullet\bullet$ & $\bullet\bullet\circ$ & $\bullet\circ\circ$ & $\bullet\bullet\bullet$ & $\bullet\circ\circ$ & $\bullet\bullet\circ$ & $\bullet\circ\circ$ & No \\
\midrule
\multicolumn{10}{l}{\emph{Family C: Project Management processes}} \\
9 & AI Use Cases~\cite{REF108} & $\bullet\bullet\circ$ & $\bullet\bullet\circ$ & $\bullet\bullet\circ$ & $\bullet\bullet\circ$ & $\bullet\circ\circ$ & $\bullet\circ\circ$ & $\bullet\circ\circ$ & No \\
10 & BDDS~\cite{REF4} & $\bullet\bullet\circ$ & $\bullet\bullet\bullet$ & $\bullet\bullet\circ$ & $\bullet\bullet\circ$ & $\bullet\circ\circ$ & $\bullet\circ\circ$ & $\bullet\circ\circ$ & No \\
11 & CRISP-DM~\cite{REF6, REF7} & $\bullet\bullet\circ$ & $\bullet\circ\circ$ & $\bullet\circ\circ$ & $\bullet\bullet\circ$ & $\bullet\circ\circ$ & $\bullet\bullet\circ$ & $\bullet\bullet\circ$ & No \\
12 & DST~\cite{REF5} & $\bullet\bullet\circ$ & $\bullet\bullet\circ$ & $\bullet\circ\circ$ & $\bullet\bullet\circ$ & $\bullet\circ\circ$ & $\bullet\bullet\circ$ & $\bullet\bullet\circ$ & No \\
13 & AI Lifecycle~\cite{REF15} & $\bullet\bullet\circ$ & $\bullet\circ\circ$ & $\bullet\circ\circ$ & $\bullet\bullet\circ$ & $\bullet\circ\circ$ & $\bullet\bullet\circ$ & $\bullet\bullet\circ$ & No \\
14 & CRISP-ML(Q)~\cite{REF17} & $\bullet\bullet\circ$ & $\bullet\circ\circ$ & $\bullet\circ\circ$ & $\bullet\bullet\bullet$ & $\bullet\bullet\bullet$ & $\bullet\bullet\bullet$ & $\bullet\bullet\bullet$ & No \\
15 & Interpret. ML~\cite{REF107} & $\bullet\bullet\bullet$ & $\bullet\circ\circ$ & $\bullet\circ\circ$ & $\bullet\bullet\circ$ & $\bullet\bullet\circ$ & $\bullet\bullet\circ$ & $\bullet\bullet\circ$ & No \\
16 & MAISTRO~\cite{REF106} & $\bullet\bullet\bullet$ & $\bullet\circ\circ$ & $\bullet\circ\circ$ & $\bullet\bullet\circ$ & $\bullet\bullet\circ$ & $\bullet\bullet\bullet$ & $\bullet\bullet\bullet$ & No \\
\midrule
\multicolumn{10}{l}{\emph{Family D: Automation-Oriented}} \\
17 & MLOps~\cite{REF19} & $\bullet\circ\circ$ & $\bullet\circ\circ$ & $\bullet\circ\circ$ & $\bullet\bullet\bullet$ & $\bullet\circ\circ$ & $\bullet\bullet\circ$ & $\bullet\bullet\bullet$ & No \\
18 & SmartML~\cite{REF18} & $\bullet\circ\circ$ & $\bullet\circ\circ$ & $\bullet\bullet\circ$ & $\bullet\circ\circ$ & $\bullet\circ\circ$ & $\bullet\circ\circ$ & $\bullet\circ\circ$ & No \\
\midrule
\multicolumn{2}{l|}{\textbf{Strong Coverage ($\bullet\bullet\bullet$)}} & 5 & 2 & \cellcolor{gray!30}0 & 5 & 1 & 2 & 3 & \textbf{0/18} \\
\multicolumn{2}{l|}{\textbf{\% Strong Coverage}} & 28\% & 11\% & \cellcolor{gray!30}0\% & 28\% & 5\% & 11\% & 17\% & \textbf{0\%} \\
\bottomrule
\end{tabular}
\begin{tablenotes}
\scriptsize
\item \textbf{Stages:} S1=Characterize Problem, S2=Formulate Decisions, S3=Select ML Task/Paradigm, S4=Specify Data Requirements, S5=Define Quality Constraints, S6=Establish Evaluation Criteria, S7=Plan Deployment \& Operations.
\item $\bullet\bullet\bullet$ = Strong (explicit guidance),  $\bullet\bullet\circ$ = Partial (some guidance),  $\bullet\circ\circ$ = Minimal (little/no guidance). Shaded column indicates the important gap.
\end{tablenotes}
\end{table*}

\paragraph{Finding 1: Incompleteness of the approaches}

All 18 approaches exhibit at least one stage with minimal coverage ($\bullet\circ\circ$). This uniform incompleteness spans all four methodological families.
The approach closest to completeness is GR4ML~\cite{REF51} with strong coverage of S1, S2, and S4, and partial coverage of S3. GR4ML provides three complementary views: Business View, Analytics Design View, and Data Preparation View linked through the central concept of \emph{Insight}, defined as the output produced by an ML model to answer a business question. Yet, industrial validation in the healthcare domain revealed that the Business View to Analytics Design View transition continues to be underspecified~\cite{REF51}. Practitioners are required to rely on intuition for paradigm selection despite the framework's otherwise rigorous structure. 

\paragraph{Finding 2: The S2--S3 Discontinuity}

The coverage matrix exhibits a discontinuity at stages S2--S3. Strong coverage drops sharply across the transformation boundary: S1 (28\%) $\rightarrow$ S2 (11\%) $\rightarrow$ S3 (0\%) $\rightarrow$ S4 (28\%). Only 2 of the 18 approaches (GR4ML, BDDS) provide strong support for S2 through explicit decision-goal constructs and Theory of Constraints (TOC)–based root-cause analysis. In contrast, the remaining 16 approaches either move directly from high-level objectives to data preparation or fail to provide a rigorous and explicit intermediate decision formulation. At S3, none of the 18 approaches provides strong guidance for ML task selection,  only four approaches (GR4ML, AI Use Cases, BDDS, and SmartML) offer partial support in this regard.

Encoding coverage as Strong=3, Partial=2, Minimal=1, the
per-stage means confirm the pattern: S1 ($\bar{x}=2.11$),
S2 ($\bar{x}=1.50$), S3 ($\bar{x}=1.22$, the lowest
of all seven stages), S4 ($\bar{x}=2.22$), S5 ($\bar{x}=1.61$),
S6 ($\bar{x}=1.83$), S7 ($\bar{x}=1.61$). The deficit is
consistent across families: mean S3 is $1.20$ for RE4AI, $1.00$
for Data-Centric, $1.25$ for PM, and $1.50$ for Automation, none
exceeds "partial" and $15$ of $18$ approaches score higher
downstream (S4--S7) than upstream (S1--S3).

\paragraph{Finding 3: Family-Level Coverage Profiles}\mbox{}
\setlength{\parskip}{0.1cm}

\textbf{Family A (RE4AI).} RE4AI approaches provide partial-to-strong coverage of S1 (problem characterization) and S4 (data specification) and perform well in cataloging AI-specific concerns and organizing requirements. However, all five offer only minimal or partial support for S3 (ML task selection). RM4ML provides no mechanism for translating business goals into task formulations~\cite{REF96}. In general, these approaches specify \emph{what} to capture, not \emph{how} to formulate the ML approach.

\textbf{Family B (Data-Centric).} 
Approaches in this family show varied S4 coverage: Multi-layer achieves strong data exploration, while Extended AMDiRE and MLC Spec. offer only partial coverage. All three score minimal on S3. Although AMDiRE supports traceability via annotations, it introduces algorithm-related artifacts without defining a systematic derivation process from business-level requirements~\cite{REF70}. Similarly, MLC Spec. and Multi-layer define data requirements based on pre-assumed tasks rather than deriving them from the characteristics of the problem~\cite{REF69, REF81}. Multi-layer starts from a fixed objective (e.g.“detect pedestrians”), where the ML decision is implicitly assumed and data are specified accordingly. Likewise, MLC Spec. begins with a predefined concept (e.g., “pedestrian”) as the MLC target, focusing on concept scoping instead of deriving the task from the problem context.


\textbf{Family C (Project Management).} Process approaches show balanced downstream coverage, with CRISP-ML(Q) achieving strong S7 through quality assurance and operational governance~\cite{REF17}. However, 6 of 8 score minimal on S3 and 5 of 8 on S2: they define when transformation occurs (phase sequencing, governance checkpoints) but do not provide guidance about how to execute it. CRISP-DM provides minimal guidance for Business Understanding despite industrial ubiquity. Partial exceptions exist: BDDS achieves strong S2 coverage via the TOC,  AI Use Cases provides partial S3 support through cognitive function mapping. Both are qualitative and offer no algorithm recommendations~\cite{REF4, REF108}.

\textbf{Family D (Automation).} Automation-oriented approaches exhibit inverted profiles, partial-to-strong downstream coverage and minimal upstream coverage. MLOps achieves S4 and S7 strong through operational principles, yet scores minimal on S1–S2, explicitly assuming that business understanding exists~\cite{REF19}. SmartML provides partial S3 coverage \emph{within constraints}: given a pre-defined classification task, it automates algorithm selection, but cannot determine whether a business problem should be classification, regression, or clustering~\cite{REF18}. Thus, both approaches begin where formulation guidance effectively ends.

\section{Synthesis: The ATP and Research Recommendations}
\label{sec:synthesis}

This section synthesizes findings from our coverage analysis. We first characterize the core transformation challenge as the Analytics Translation Problem (Section~\ref{subsec:atp}), and then derive recommendations from the identified gaps~(Section~\ref{subsec:recommendations}).

\subsection{The Analytics Translation Problem (ATP)}
\label{subsec:atp}

\begin{figure}[htbp]
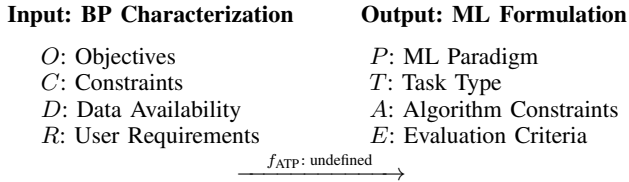

\centering
\small
\begin{tabular}{c@{\hspace{0.8cm}}c}
\textbf{Input: BP Characterization} & \textbf{Output: ML Formulation} \\[6pt]
\begin{tabular}{l}
$O$: Objectives \\
$C$: Constraints \\
$D$: Data Availability \\
$R$: User Requirements
\end{tabular}
&
\begin{tabular}{l}
$P$: ML Paradigm \\
$T$: Task Type \\
$A$: Algorithm Constraints \\
$E$: Evaluation Criteria
\end{tabular}
\\[10pt]
\multicolumn{2}{c}{$\xrightarrow{\quad f_{\text{ATP}}: \text{ undefined} \quad}$}
\end{tabular}
\caption{The Analytics Translation Problem: the S2--S3 formulation gap.}
\label{fig:atp}
\end{figure}

The capability gaps shown by our coverage analysis converge on a single challenge, which we characterize as the \emph{Analytics Translation Problem} (ATP). Figure~\ref{fig:atp} frames the ATP as a problem statement rather than a formal specification: it identifies the input and output component spaces that a future translation function must relate, without prescribing the function's form. Characterizing \emph{what} must be mapped is a prerequisite for formalizing \emph{how},  the latter requires the empirical work outlined in our future directions.

Table~\ref{tab:atp-mapping} shows the relationship between ATP components and the artifact categories from Tables~\ref{tab:input_artifacts}--\ref{tab:output_artifacts}. On the input side, Strategic Objectives and Stakeholder Specification jointly define what the organization needs ($O$, $R$),  Constraints and Rules define the boundaries within which any solution must operate ($C$),  and Data Characteristics determine which learning approaches are feasible ($D$). Domain Context is excluded because it provides the frame within which inputs are understood but is not itself transformed into an ML construct. Feasibility Assessment is excluded because it operates as a validation gate rather than a translation input. On the output side, Algorithm/Task Specification maps to the core formulation decisions ($P$, $T$, $A$) and Evaluation Criteria to success metrics ($E$). Other outputs, Data Requirements, Deployment Considerations, and Traceability Links, are derived from the formulation and do not constitute it.

\begin{table}[htbp]
\centering
\caption{Mapping Between Artifact Categories and ATP Components}
\label{tab:atp-mapping}
\scriptsize
\renewcommand{\arraystretch}{1.15}
\begin{tabular}{p{2.8cm}p{1.0cm}p{3.5cm}}
\toprule
\textbf{Artifact Category} & \textbf{ATP} & \textbf{Role} \\
\midrule
\multicolumn{3}{l}{\emph{Input (Table~\ref{tab:input_artifacts})}} \\
Strategic Objectives & $O$ & Goals and decision needs \\
Stakeholder Specification & $R$ & Trust and control expectations \\
Constraints \& Rules & $C$ & Solution boundaries \\
Data Characteristics & $D$ & Feasible learning approaches \\
Domain Context & --- & Interpretive frame (not transformed) \\
Feasibility Assessment & --- & Validation gate (not translated) \\
\midrule
\multicolumn{3}{l}{\emph{Output (Table~\ref{tab:output_artifacts})}} \\
Algorithm/Task Spec. & $P, T, A$ & Core formulation decisions \\
Evaluation Criteria & $E$ & Success metrics aligned with $O$ \\
Data Requirements & --- & Derived from selected $P, T$ \\
Quality Attr. / NFRs & --- & Input via $C$; constrains $A$ \\
Deployment Considerations & --- & Post-formulation (S7) \\
Traceability Links & --- & Cross-cutting quality (R4) \\
\bottomrule
\end{tabular}
\end{table}

Our coverage analysis shows that approaches model input and output components individually,  but none provides the translation function $f_{\text{ATP}}$ mapping $(O, C, D, R) \rightarrow (P, T, A, E)$. The following recommendations target the specific gaps that prevent this translation.

\subsection{Research Recommendations}
\label{subsec:recommendations}

Each recommendation draws on evidence from the corpus and includes an illustrative example. We stress that these examples sketch the shape of a solution, not its final form, the specific thresholds, ratings, and formulations would need calibration through expert input and industrial case studies before anyone should adopt them. Operationalizing the recommendations into a working framework is a future work.

\subsubsection{R1: Support Multi-Formulation Exploration}

\emph{Gap.} A single business problem admits multiple valid ML formulations classification versus regression, supervised versus unsupervised each with different data requirements, feasibility constraints, and trade-offs. 
GR4ML~\cite{REF51} provides a multi-view representation, AI Use Cases~\cite{REF108} maps business capabilities to multiple AI solution types, and BDDS~\cite{REF4} uses root cause analysis to explore alternative framings. Yet, all treat views as sequential refinements toward a single solution, rather than as parallel alternatives for systematic comparison.

\emph{Recommendation.} Future frameworks should extend GR4ML's multi-view structure with an explicit \emph{formulation space} representation. While GR4ML transitions linearly from the Business View to the Analytics Design View, an extended approach would maintain alternative formulations as artifacts throughout specification, with structured comparison criteria including data requirements, feasibility indicators, business alignment, and implementation cost.

\emph{Example.} Given objective ``reduce customer churn,'' a formulation space might contain:

\begin{center}
\small
\begin{tabular}{p{2.5cm} p{1.7cm} p{0.7cm} p{2.4cm}}
\emph{Formulation} & \emph{Data Req.} & \emph{Feasib.} & \emph{Trade-off} \\ \hline
Predict who churns & Churn labels & High & Direct but reactive \\
Predict timing & Temporal data & Medium & Enables proaction \\
Identify segments & No labels & High & Less precise \\
Recommend actions & Feedback loop & Low & Optimizes retention \\
\end{tabular}
\end{center} 

\subsubsection{R2: Provide Task Derivation Guidance}

\emph{Gap.} The S2$\rightarrow$S3 translation consists of the transition from decision questions to ML task specification. Practitioners independently determine whether a problem requires classification, regression, clustering, or other tasks, relying on expertise rather than systematic guidance. GR4ML~\cite{REF51} decomposes goals into question goals but provides no formal guidance for selecting which ML task produces the required insight,  question goals may be vague, and the approach lacks mechanisms for prioritization or alignment with business value. BDDS~\cite{REF4} adds the TOC for focused decomposition but addresses root causes rather than task selection. AI Use Cases~\cite{REF108} maps seven cognitive functions to problem types but provides coarse categories without decision rules. Data-Centric approaches~\cite{REF70, REF69, REF81} encode implicit paradigm selection based on data characteristics but leave this logic undocumented. The question ``should this be classification or clustering?'' receives no systematic support.

\emph{Recommendation.} Future frameworks should combine GR4ML's goal decomposition with explicit derivation rules, leveraging AI Use Cases' cognitive mapping and making explicit what Data-Centric approaches leave implicit. Task-type selection should be supported through two rule categories: \emph{structural rules}, linking question types to task categories based on what the question asks, and \emph{data rules}, filtering candidates according to available data characteristics.

\emph{Example.} Derivation rules and data feasibility filters:

\begin{center}
\small
\begin{tabular}{p{3.5cm} p{3cm}}
\emph{If data shows...} & \emph{Then exclude...} \\ \hline
Labels unavailable & Supervised tasks \\
No temporal structure & Forecasting, survival \\
No feedback mechanism & Reinforcement learning \\
\end{tabular}
\end{center}

\begin{center}
\small
\begin{tabular}{p{4.5cm} p{3.5cm}}
\emph{If question asks...} & \emph{Then candidate task...} \\ \hline
Which category does X belong to? & Classification \\
What is the value of Y? & Regression \\
When will event Z occur? & Survival, forecasting \\
What groups exist? & Clustering \\
Is this observation unusual? & Anomaly detection \\
What action maximizes outcome? & Reinforcement learning \\
\end{tabular}
\end{center}

\subsubsection{R3: Enable Constraint-Algorithm Filtering}

\emph{Gap.} Constraints such as explainability, fairness, latency, and interpretability are captured as requirements but rarely translated into algorithm selection guidance. RE4HCAI~\cite{REF67} and the Catalogue of Concerns~\cite{REF71} provide constraint taxonomies,  Extended AMDiRE~\cite{REF70} structures constraints within artifact models,  and CRISP-ML(Q)~\cite{REF17} incorporates constraints into model selection by narrowing the set of candidate models against quality dimensions, but without specifying how individual constraints exclude particular algorithm families. Most approaches catalogue \emph{what} constraints exist without specifying \emph{how} they restrict algorithm choice. Constraint verification occurs post-hoc after algorithm selection rather than guiding it. If a real-time application requires sub-second latency and auditable fairness, no systematic guidance exists for determining compatible algorithm families.

\emph{Recommendation.} Future frameworks should extend RE4HCAI's constraint taxonomy with \emph{compatibility matrices} mapping constraint types to algorithm family suitability. Rather than post-hoc verification, an extended approach would enable pre-selection filtering through eliminative reasoning: constraints progressively exclude incompatible algorithm families, yielding candidates that satisfy all requirements before detailed selection begins.

\emph{Example.} Given $C$ = \{latency $<$100ms, fairness auditable\}, compatibility filtering yields:

\begin{center}
\small
\begin{tabular}{p{3.2cm} c c c}
\emph{Constraint} & \emph{Neural} & \emph{Ensemble} & \emph{Linear} \\ \hline
Latency $<$100ms & \ding{55} & \ding{51} & \ding{51} \\
Fairness auditable & Partial & Partial & \ding{51} \\
High accuracy important & \ding{51} & \ding{51} & Partial \\
\end{tabular}
\end{center}

\subsubsection{R4: Implement Probabilistic Traceability}

\emph{Gap.} Traditional traceability assumes binary satisfaction, a requirement is either met or not. ML solutions, still, partially satisfy objectives with quantifiable uncertainty: a classifier achieving 85\% accuracy addresses the goal imperfectly but potentially usefully. Stakeholders cannot assess specification adequacy without understanding expected performance ranges. GR4ML~\cite{REF51} introduces the Insight concept linking business questions to analytics outputs, but provides structural links without expected performance. Multi-layered Framework~\cite{REF81} quantifies uncertainty, but only for data fitness rather than for business-objective satisfaction. CRISP-ML(Q)~\cite{REF17} maintains quality documentation but does not propagate implications to business objectives. Most of the approaches lack explicit traceability from business objectives through requirements to ML decisions,  when models underperform, tracing which objective is affected becomes impossible. 

\emph{Recommendation.} Future frameworks should extend GR4ML's Insight-based traceability with probabilistic annotations: (1)~expected performance ranges based on domain benchmarks or preliminary analysis, (2)~confidence levels indicating estimated reliability, and (3)~business impact translations specifying what ranges mean for objective satisfaction. Such annotations would propagate through the traceability chain, transforming it from documentation into actionable decision support.

\emph{Example.} Annotated trace from business goal to  metric:
\begin{center}
\small
\begin{tabular}{p{8.5cm}}
\textbf{Goal:} Reduce fraud losses by 20\% \\
\hspace{0.5cm}$\downarrow$ \emph{[expected: 12--25\%,  confidence: medium]} \\
\textbf{Decision:} Flag fraudulent transactions for review \\
\hspace{0.5cm}$\downarrow$ \emph{[expected: 80\% caught, 5\% false positive]} \\
\textbf{Task:} Binary classification with threshold \\
\hspace{0.5cm}$\downarrow$ \emph{[basis: 50K samples, domain benchmarks]} \\
\textbf{Metric:} Precision $\geq$0.85 at recall $\geq$0.80 \\
\end{tabular}
\end{center}

\subsubsection{R5: Define Data-Triggered Revision Paths}

\emph{Gap.} Formulation feasibility depends on data characteristics discoverable only through exploration label quality, feature-target relationships, and distributional properties. Unlike traditional RE, where iteration is stakeholder-triggered, BP$\rightarrow$MLS requires data-triggered revision where empirical findings invalidate upstream decisions. CRISP-DM~\cite{REF6, REF7} acknowledges iteration through circular process arrows but without specifying conditions. CRISP-ML(Q)~\cite{REF17} introduces quality checkpoints but verifies quality without defining revision triggers. AI Lifecycle revisions~\cite{REF15} add Go/No-Go gates but make binary proceed/stop decisions rather than indicating which upstream decision to revise. MAISTRO~\cite{REF106} demonstrates ethics-triggered revision but does not generalize to data findings. Without explicit triggers, teams may abandon promising directions prematurely or persist with infeasible formulations. 

\emph{Recommendation.} Future frameworks should extend CRISP-ML(Q)'s checkpoint mechanism with explicit \emph{revision triggers} and \emph{revision paths}: (1)~diagnostic conditions under which data findings mandate reconsideration, and (2)~targeted paths indicating which specific upstream recommendation (R1, R2, or R3) to revisit. Building on MAISTRO's ethics triggers, such mechanisms could generalize to data-quality, feasibility, and bias conditions.

\emph{Example.} Revision trigger structure:
\begin{center}
\small
\begin{tabular}{p{2.5cm} p{2cm} p{3.2cm}}
\emph{Data finding} & \emph{Revise} & \emph{Alternative path} \\ \hline
Labels $<$ threshold & R2 (paradigm) & Consider unsupervised \\
Imbalance $>$ ratio & R2 (task) & Anomaly detection \\
Signal $<$ threshold & R1 (objective) & May be infeasible \\
Bias detected & R3 (constr.) & Add fairness requirement \\
\end{tabular}
\end{center}

\subsubsection{Integration}

The five recommendations can work as a pipeline in practice. The process begins with R1: rather than committing to a single ML formulation, the team lays out alternatives for the same business problem. R2 populates that space—derivation rules match question types to candidate tasks and eliminate those lacking suitable data. R3 then narrows the field further, filtering out candidates that violate non-functional requirements like latency or fairness. With a shortlist in hand, stakeholders need a basis for comparison. R4 provides this by attaching expected performance ranges and tracing them back to business objectives. Finally, R5 defines when to revisit earlier decisions: if data exploration shows the label scarcity, severe class imbalance, or weak signal, the team returns to R1, R2, or R3 rather than forcing a formulation that the data cannot support. Taken together, the recommendations specify what a future framework would need to do : generate alternatives, derive tasks, filter by constraints, inform selection, and trigger revision. Our five recommendations thus define a research agenda not only for methodological frameworks but for the design of effective human-AI collaboration in an important phase of AI projects.

\section{Discussion}
\label{sec:discussion}

\subsection{Implications for Requirements Engineering}
\label{subsec:impl-re}
The upstream transformation problem, i.e., translating business characterizations into ML formulations, remains underaddressed even in approaches aiming to bring RE rigor to AI development~\cite{REF68, REF60}. Existing RE4AI research has made progress in listing AI-specific concerns~\cite{REF71, REF67} and structuring non-functional requirements~\cite{REF62, REF96}. However, these contributions predominantly focus on the solution space: specifying what an AI system should satisfy once the ML task is known~\cite{REF59, REF70}. The step of deciding which ML task to formulate remains outside the scope of all five RE4AI approaches we reviewed. We argue that established RE concepts are suitable for constructing the ATP if they are extended to support cross-domain derivation. Goal decomposition~\cite{REF20, REF21}, illustrated by GR4ML's three-view structure~\cite{REF51}, provides a structured pathway for refinement from business objectives to technical formulations. Quality requirement frameworks~\cite{iso25010, REF62} can model constraint–algorithm compatibility relations, as articulated in R3. Traceability mechanisms~\cite{REF70, REF81}, extended with the probabilistic annotations proposed in R4, can preserve the business-to-technical reasoning chain that current approaches lack. The RE constructs exist; the hard part is adapting them to work across the business-to-ML boundary.

\subsection{Implications for Practice}
\label{subsec:impl-practice}

We argue that practitioners should treat ML task formulation as an explicit requirements activity rather than an implicit data science decision. Our stage coverage analysis shows that the S2–S3 transition is where methodological support is weakest. To reduce the misalignment identified in~\cite{Ryseff2024} practitioners should make this step explicit with documented alternatives, clear selection rationale, and stakeholder validation.
The input and output artifact categories in Tables~\ref{tab:input_artifacts} and~\ref{tab:output_artifacts} can serve as a completeness checklist. Our analysis indicates that no approach covers all six input categories. This means that practitioners who rely on only one methodology are likely to encounter blind spots. Cross-referencing artifacts across different families—for example, combining BDDS’s root cause analysis with RE4HCAI’s constraint taxonomy and a Data-Centric approach’s data quality criteria can provide broader coverage than any single approach alone. 
Our findings show that the BP→MLS transformation requires iteration based on empirical data results, not solely on stakeholder feedback. Therefore, project plans should include explicit decision points where insights from data exploration can prompt revisiting earlier choices, without treating such revision as a project failure.

\subsection{Threats to Validity}
\label{subsec:threats}

\textbf{Internal validity.} Extraction and classification were primarily conducted by the first author, with co-authors review and discussions on limited to ambiguous cases. Inter-rater reliability was assessed on a 30\% sample ($\kappa = 0.83$,  see Section~\ref{subsec:analysis-framework}). The three-level rating scale involves subjective judgment, which we mitigated through explicit criteria (Table~\ref{tab:criteria}) and the public availability of all extracted data.

\textbf{External validity.} Our corpus of 18 approaches may not include unpublished industry practices or work published outside our search scope. We mitigated selection bias by searching five digital libraries, targeting key RE and AI venues, and applying a theoretical saturation stopping rule. That said, approaches published after our search period are not covered.

\textbf{Construct validity.} The seven-stage framework (S1–S7) is deliberately derived from the corpus rather than imposed externally. This way, the gaps we find are gaps in current practice and not assumptions external to the reviewed approaches. We limited the resulting circularity by grounding stages externally (Section~\ref{subsec:coverage}) and by interpreting gaps as inconsistencies revealed by the approaches. The 18 approaches themselves consider S2–S3 as necessary steps, yet none offers concrete mechanisms for carrying them out. We interpret this as a recognized inconsistency not a failure measured against an external benchmark. Validation of the framework on real project workflows is left for future work. The four-family taxonomy captures dominant methodological orientations, although some approaches could be classified differently under alternative criteria.

\section{Related Work}
\label{sec:related}

Recent systematic reviews converge on the same conclusion about RE4AI. Habiba et al.~\cite{REF111} found that analysis and elicitation dominate (104 and 87 of 126 studies), while process support continues to be marginal. Ahmad et al.~\cite{REF68, REF60} reached a similar conclusion: among 43 studies, only three frameworks addressed elicitation-level concerns, and none guided ML task determination. Villamizar et al.~\cite{REF105} and Nasri et al.~\cite{REF110}  confirmed that existing approaches primarily operate after the ML task has been decided, with specification accounting for 62\% of contributions and validation receiving far less attention.The pattern extends beyond RE. Shimaoka et al.~\cite{REF113} reviewed 17 CRISP-DM adaptations and found that none restructures Business Understanding to formalize task derivation, even upstream variants assume the learning task is predefined. Alcoba\c{c}a and de Carvalho~\cite{REF112} synthesized 52 AutoML surveys and observed the same boundary: AutoML automates modeling under a given task but does not address how that task emerges from a business problem. Across all three streams, RE4AI, data science processes, and AutoML, ML task derivation is  largely implicit. Our work addresses the prior step. Where existing reviews deal with the specification of requirements for AI systems, in this paper, we focus on deriving the ML task itself  from the business problem.

\section{Conclusion and Future Work}
\label{sec:conclusion}

This paper examines how 18 approaches from requirements engineering, ML project management, and automation support the translation from business problems to ML solution specifications. Most approaches specify what outputs an ML project should produce. However, they provide little guidance on how to derive them from business characterizations. The upstream stages, decision formulation and ML task selection, have not yet been supported across all four families, even though the approaches themselves acknowledge these stages as necessary. We characterize this gap as the Analytics Translation Problem (ATP). ATP is the absence of  operationalizable mappings between business intents and ML formulations. We derive five research recommendations (R1--R5) that define what a future transformation framework would need to provide. The building blocks are already present across the reviewed approaches (e.g., goal decomposition, constraint taxonomies, quality checkpoints). However, no approach assembles them into a complete pipeline. Our recommendations identify where these pieces exist and what is still missing.

Several directions are open. First, the ATP formalization and proposed recommendations require empirical validation through industrial case studies to assess their practical utility. Second, the seven-stage framework should be evaluated against real project workflows to confirm that it captures the activities practitioners perform. Third, the illustrative examples accompanying R1--R5 need calibration via domain expert elicitation and benchmark studies before they can serve as operational guidance. This work reframes the BP→MLS transformation as a distinct cross-domain refinement problem within RE. In addition, we establish a conceptual and analytical foundation for developing principled, traceable mechanisms to replace the current reliance on practitioner intuition. These findings are especially timely as organizations integrate generative AI assistants into their workflows. Thus, formalizing the ATP is a prerequisite for AI tools that can meaningfully assist this translation.

\section*{Data Availability Statement}
We provide the replication package for this study, containing the literature review dataset (BP2AIS) on Zenodo\footnote{\url{https://doi.org/10.5281/zenodo.18779531}}.

\bibliographystyle{IEEEtran}
\bibliography{references}

\end{document}